\documentclass[12pt]{article} \usepackage[journal]{optional}

\usepackage{amsmath}
\usepackage{amssymb}
\usepackage{version}
\usepackage{enumerate}

\opt{journal}{
    \includeversion{journal}
    \excludeversion{conference-with-appendix}
    \excludeversion{conference-no-appendix}
}
\opt{conference-with-appendix}{
    \includeversion{conference-with-appendix}
    \excludeversion{conference-no-appendix}
    \excludeversion{journal}
}
\opt{conference-no-appendix}{
    \includeversion{conference-no-appendix}
    \excludeversion{conference-with-appendix}
    \excludeversion{journal}
}

\newcommand{\qedllncs}{
    \opt{conference-with-appendix,conference-no-appendix}{\qed}
}
\newcommand{\ve}[1]{\overrightarrow{#1}}
\newcommand{\ma}[1]{#1}
\newcommand{\major}{\succcurlyeq}

\date{}
\title{{\bf Finite-State Dimension and Real Arithmetic}}

\opt{journal}{
    \oddsidemargin  0.0in
\evensidemargin 0.0in
\textwidth      6.3in
\headheight     0.0in
\topmargin      -0.3in
\textheight     9.0in

    \usepackage{commands}
    \usepackage{numbering}
    \author{
        David Doty \thanks{Department of Computer Science, Iowa State University, Ames, IA 50011 USA. ddoty {\it at} iastate {\it dot} edu. This research was funded in part by grant number 9972653 from the National Science Foundation as part of their Integrative Graduate Education and Research Traineeship (IGERT) program.}
        \and
        Jack H. Lutz \thanks{Department of Computer Science, Iowa State University, Ames, IA 50011 USA. lutz {\it at} cs {\it dot} iastate {\it dot} edu. This research was supported  in part by National Science Foundation Grant 0344187.}
        \and
        Satyadev Nandakumar \thanks{Department of Computer Science, Iowa State University, Ames, IA 50011 USA. satyadev {\it at} iastate {\it dot} edu. This research was supported  in part by National Science Foundation Grant 0344187.}
    }
}

\opt{conference-with-appendix,conference-no-appendix}{
    \usepackage{commands}
    \numberwithin{equation}{section}
    \numberwithin{theorem}{section}
    \spnewtheorem{cor}[theorem]{Corollary}{\bfseries}{\itshape}
    \spnewtheorem{lem}[theorem]{Lemma}{\bfseries}{\itshape}
    \spnewtheorem{obs}[theorem]{Observation}{\bfseries}{\itshape}
    \spnewtheorem{defn}{Definition}{\bfseries}{\itshape}
    \spnewtheorem*{ack}{Acknowledgments}{\bfseries}{\rmfamily}

    \author{
        David Doty \inst{}
        \thanks{This research was supported in part by National Science Foundation Grant 9972653 as part of their Integrative Graduate Education and Research Traineeship (IGERT) program.}
        \and
        Jack H. Lutz \inst{}
        \thanks{This research was supported  in part by National Science Foundation Grant 0344187.}
        \and
        Satyadev Nandakumar \inst{}
        \thanks{This research was supported  in part by National Science Foundation Grant 0344187.}
    }
    \institute{ Department of Computer Science, Iowa State University, Ames, IA 50011, USA. \\ \email{\{ddoty,lutz,satyadev\} {\it at} cs {\it dot} iastate {\it dot} edu}}
}

\begin{document}
\maketitle
\begin{abstract}
We use entropy rates and Schur concavity to prove that, for every integer $k \geq 2$, every nonzero rational number $q$, and every real number $\alpha$, the base-$k$ expansions of $\alpha, q+\alpha,$ and $q \alpha$ all have the same finite-state dimension and the same finite-state strong dimension.  This extends, and gives a new proof of, Wall's 1949 theorem stating that the sum or product of a nonzero rational number and a Borel normal number is always Borel normal.
\end{abstract}

\section{Introduction}
The finite-state dimension of a sequence $S$ over a finite alphabet $\Sigma$ is an asymptotic measure of the density of information in $S$ as perceived by finite-state automata.  This quantity, denoted $\dimfs(S)$, is a finite-state effectivization of classical Hausdorff dimension \cite{Haus19,Falc90} introduced by Dai, Lathrop, Lutz, and Mayordomo \cite{Dai:FSD}.  A dual quantity, the finite-state \emph{strong} dimension of $S$, denoted $\Dimfs(S)$, is a finite-state effectivization of classical packing dimension \cite{Tricot82,Sull84,Falc90} introduced by Athreya, Hitchcock, Lutz, and Mayordomo \cite{Athreya:ESDAICC}. (Explicit definitions of $\dimfs(S)$ and $\Dimfs(S)$ appear in section \ref{sec-prelim}.)  In fact both $\dimfs(S)$ and $\Dimfs(S)$ are asymptotic measures of the density of finite-state information in $S$, with $0 \leq \dimfs(S) \leq \Dimfs(S) \leq 1$ holding in general and $\dimfs(S) = \Dimfs(S)$ holding when $S$ is sufficiently ``regular.''

Although finite-state dimension and finite-state strong dimension were originally defined in terms of finite-state gamblers \cite{Dai:FSD,Athreya:ESDAICC} (following the gambling approach used in the first effectivizations of classical fractal dimension \cite{Lutz:DCC,Lutz:DISS}), they have also been shown to admit equivalent definitions in terms of information-lossless finite-state compressors \cite{Dai:FSD,Athreya:ESDAICC}, finite-state predictors in the log-loss model \cite{Hitchcock:FDLLU,Athreya:ESDAICC}, and block-entropy rates \cite{Bourke:ERFSD}.  In each case, the definitions of $\dimfs(S)$ and $\Dimfs(S)$ are exactly dual, differing only that a limit inferior appears in one definition where a limit superior appears in the other.  These two finite-state dimensions are thus, like their counterparts in fractal geometry, robust quantities and not artifacts of a particular definition.

The sequences $S$ satisfying $\dimfs(S) = 1$ are precisely the \emph{(Borel) normal} sequences, i.e., those sequences in which each nonempty string $w \in \Sigma^*$ appears with limiting frequency $|\Sigma|^{-|w|}$.  (This fact was implicit in the work of Schnorr and Stimm \cite{SchSti72} and pointed out explicitly in \cite{Bourke:ERFSD}.)  The normal sequences, introduced by Borel in 1909 \cite{Bore09}, were extensively investigated in the twentieth century \cite{Nive56,KuiNie74,Weis99,DajKra02,Harm03}. Intuitively, the normal sequences are those sequences that are random relative to finite-state automata. This statement may seem objectionable when one first learns that the Champernowne sequence
$$
    0100011011000001010011100 \ldots,
$$
obtained by concatenating all binary strings in standard order, is normal \cite{Cham33}, but it should be noted that a finite-state automaton scanning this sequence will spend nearly all its time in the middle of long strings that are random in the (stronger) sense of Kolmogorov complexity \cite{LiVi97} and, having only finite memory, will have no way of ``knowing'' where such strings begin or end.  This perspective is especially appropriate when modeling situations in which a data stream is truly massive relative to the computational resources of the entity processing it.

An informative line of research on normal sequences concerns operations that preserve normality.  For example, in his 1949 Ph.D. thesis under D.H. Lehmer, Wall \cite{Wall49} proved that every subsequence that is selected from a normal sequence by taking all symbols at positions occurring in a given arithmetical progression is itself normal. Agafonov \cite{Agaf68} extended this by showing that every subsequence of a normal sequence that is selected using a regular language is itself normal; Kamae \cite{Kam73} and Kamae and Weiss \cite{KamWei75} proved related results; and Merkle and Reimann \cite{MerRei03} proved that a subsequence selected from a normal sequence using a context-free language need not be normal (in fact, can be constant, even if selected by a one-counter language).  For another example, again in his thesis, Wall \cite{Wall49} (see also \cite{KuiNie74,BorBai04}) proved that, for every integer $k \geq 2$, every nonzero rational number $q$, and every real number $\alpha$ that is normal base $k$ (i.e., has a base-$k$ expansion that is a normal sequence), the sum $q+\alpha$ and the product $q \alpha$ are also normal base $k$.  (It should be noted that a real number $\alpha$ may be normal in one base but not in another \cite{Cass59,Schm60}.)

This paper initiates the study of operations that preserve finite-state dimension and finite-state strong dimension.  This study is related to, but distinct from, the study of operations that preserve normality.  It is clear that every operation that preserves finite-state dimension must also preserve normality, but the converse does not hold.  For example, a subsequence selected from a sequence according an arithmetical progression need not have the same finite-state dimension as the original sequence. This is because a sequence with finite-state dimension less than 1 may have its information content distributed heterogeneously.  Specifically, given a normal sequence $S$ over the alphabet $\{0,1\}$, define a sequence $T$ whose $\nth$ bit is the ${\frac{n}{2}}^{\rm th}$ bit of $S$ if $n$ is even and 0 otherwise.  Then the sequence $S$ and the constant sequence $0^\infty$ are both selected from $T$ according to arithmetic progressions, but it is easy to verify that $\dimfs(T) = \Dimfs(T) = \frac{1}{2}, \dimfs(0^\infty) = \Dimfs(0^\infty) = 0$, and $\dimfs(S) = \Dimfs(S) = 1$.  Hence, Wall's first above-mentioned theorem does not extend to the preservation of finite-state dimension.  Of course, this holds \emph{a fortiori} for the stronger results by Agafonov, Kamae, and Weiss.

Our main theorem states that Wall's second above-mentioned theorem, unlike the first one, does extend to the preservation of finite-state dimension. That is, we prove that, for every integer $k \geq 2$, every nonzero rational number $q$, and every real number $\alpha$, the base-$k$ expansions of $\alpha, q+\alpha,$ and $q \alpha$ all have the same finite-state dimension and the same finite-state strong dimension.

The proof of our main theorem does not, and probably cannot, resemble Wall's uniform distribution argument.  Instead we use Bourke, Hitchcock, and Vinodchandran's block-entropy rate characterizations of $\dimfs$ and $\Dimfs$ \cite{Bourke:ERFSD}, coupled with the Schur concavity of the entropy function \cite{Schu23,MarOlk79,Bhat97}, to prove that finite-state dimension and finite-state strong dimension are contractive functions with respect to a certain ``logarithmic block dispersion'' pseudometric that we define on the set of all infinite $k$-ary sequences.  (A function is contractive if the distance between its values at sequences $S$ and $T$ is no more than the pseudodistance between $S$ and $T$.)  This gives a general method for bounding the difference between the finite-state dimensions, and the finite-state strong dimensions, of two sequences.  We then use this method to prove our main theorem.  In particular, this gives a new proof of Wall's theorem on the sums and products of rational numbers with normal numbers.

In summary, our main result is a fundamental theorem on finite-state dimension that is a quantitative extension of a classical theorem on normal numbers but requires a different, more powerful proof technique than the classical theorem.

\section{Preliminaries} \label{sec-prelim}

Throughout this paper, $\Sigma = \{0,1,\ldots,k-1\}$, where $k \geq 2$ is an integer. All \emph{strings} are elements of $\Sigma^*$, and all \emph{sequences} are elements of $\Sigma^\infty$. If $x$ is a string or sequence and $i,j$ are integers, $x[i \twodots j]$ denotes the string consisting of the $i\Th$ through $j\Th$ symbols in $x$, provided that these symbols exist. We write $x[i] = x[i \twodots i]$ for the $i\Th$ symbol in $x$, noting that $x[0]$ is the leftmost symbol in $x$. If $w$ is a string and $x$ is a string or sequence, we write $w \sqsubseteq x$ to indicate that $w = x[0 \twodots n-1]$ for some nonnegative integer $n$.

A \emph{base-$k$ expansion} of a real number $\alpha \in [0,1]$ is a sequence $S \in \Sigma^\infty$ such that
$$
    \alpha = \sum_{n=0}^\infty S[n] k^{-(n+1)}.
$$
A sequence $S \in \Sigma^\infty$ is \emph{(Borel) normal} if, for every nonempty string $w\in\Sigma^+$
$$
    \limn \frac{1}{n} \left| \left\{ \left. u \in \Sigma^{<n} \right| uw \sqsubseteq S \right\} \right| = |\Sigma|^{-|w|},
$$
i.e., if each string $w$ appears with asymptotic frequency $k^{-|w|}$ in $S$.

If $\Omega$ is a nonempty finite set, we write $\Delta(\Omega)$ for the set of all (discrete) probability measures on $\Omega$, i.e., all functions $\pi: \Omega \to [0,1]$ satisfying $\sum\limits_{w \in \Omega} \pi(w)$ $= 1$. We write $\Delta_n = \Delta(\{ 1,\ldots,n \})$.

All logarithms in this paper are base 2. The \emph{Shannon entropy} of a probability measure $\pi\in\Delta(\Omega)$ is
$$
    H(\pi) = \sum_{w\in\Omega} \pi(w) \log \frac{1}{\pi(w)},
$$
where $0 \log \frac{1}{0} = 0$.

We briefly define finite-state dimension and finite-state strong dimension. As noted in the introduction, several equivalent definitions of these dimensions are now known. In this paper, it is most convenient to use the definitions in terms of block-entropy rates, keeping in mind that Bourke, Hitchcock, and Vinodchandran \cite{Bourke:ERFSD} proved that these definitions are equivalent to earlier ones.

For nonempty strings $w,x \in \Sigma^+$, we write
$$
    \#_{\Box}(w,x) = \left| \left\{ \left. m \leq \frac{|x|}{|w|} - 1 \ \right|\ x[m |w| \twodots (m+1) |w| - 1]\right\} \right|
$$
for the number of \emph{block occurrences} of $w$ in $x$. Note that $0 \leq \#_{\Box}(w,x) \leq \frac{|x|}{|w|}$.

For each sequence $S\in\Sigma^\infty$, positive integer $n$, and string $w\in\Sigma^{<n}$, the $\nth$ \emph{block frequency} of $w$ in $S$ is
$$
    \pi_{S,n}(w) = \frac{\#_\Box(w,S[0 \twodots n|w| - 1])}{n}.
$$
Note that, for all $S\in\Sigma^\infty$ and $0<l<n$,
$$
    \sum_{w\in\Sigma^l} \pi_{S,n}(w) = 1,
$$
i.e., $\pi_{S,n}^{(l)} \in \Delta(\Sigma^l)$, where we write $\pi_{S,n}^{(l)}$ for the restriction of $\pi_{S,n}$ to $\Sigma^l$.

For each sequence $S\in\Sigma^\infty$ and positive integer $l$, the $l\Th$ \emph{normalized lower and upper block entropy rates} of $S$ are
$$
    H_l^-(S) = \frac{1}{l \log k} \liminfn H\left( \pi_{S,n}^{(l)} \right)
$$
and
$$
    H_l^+(S) = \frac{1}{l \log k} \limsupn H\left( \pi_{S,n}^{(l)} \right),
$$
respectively.
\begin{defn}
Let $S\in\Sigma^\infty$.
\begin{enumerate}
    \item The \emph{finite-state dimension} of $S$ is
    $$
        \dimfs(S) = \inf_{l\in\Z^+} H_l^-(S).
    $$
    \item The \emph{finite-state strong dimension} of $S$ is
    $$
        \Dimfs(S) = \inf_{l\in\Z^+} H_l^+(S).
    $$
\end{enumerate}
\end{defn}
More discussion and properties of these dimensions appear in the references cited in the introduction, but this material is not needed to follow the technical arguments in the present paper.

\section{Logarithmic Dispersion and Finite-State Dimension} \label{sec-logdis-fsd}
In this section we prove a general theorem stating that the difference between two sequences' finite-state dimensions (or finite-state strong dimensions) is bounded by a certain ``pseudodistance'' between the sequences. Recall that $\Delta_n = \Delta(\{1,\ldots,n\})$ is the set of all probability measures on $\{1,\ldots,n\}$.
\begin{defn} \label{def-log-disp}
Let $n$ be a positive integer. The \emph{logarithmic dispersion} (briefly, the \emph{log-dispersion}) between two probability measures $\pi,\mu\in\Delta_n$ is
$$
    \delta(\pi,\mu) = \log m,
$$
where $m$ is the least positive integer for which there is an $n \times n$ nonnegative real matrix $\ma{A}=(a_{ij})$ with the following three properties.
\begin{enumerate}[(i)]
    \item $\ma{A}$ is \emph{stochastic}: each column of $\ma{A}$ sums to 1, i.e., $\sum_{i=1}^n a_{ij} = 1$ holds for all $1 \leq j \leq n$.
    \label{log-disp-prop-1}
    \item $\ma{A} \pi = \mu$, i.e., $\sum_{j=1}^n a_{ij} \pi(j) = \mu(i)$ holds for all $1 \leq i \leq n$.
    \label{log-disp-prop-2}
    \item No row or column of $\ma{A}$ contains more than $m$ nonzero entries.
    \label{log-disp-prop-3}
\end{enumerate}
\end{defn}
It is clear that $\delta: \Delta_n \times \Delta_n \to [0,\log n]$. We now extend $\delta$ to a normalized function $\delta^+: \Sigma^\infty \times \Sigma^\infty \to [0,1]$. Recall the block-frequency functions $\pi_{S,n}^{(l)}$ defined in section \ref{sec-prelim}.
\begin{defn}
The \emph{normalized upper logarithmic block dispersion} between two sequences $S,T\in\Sigma^\infty$ is
$$
    \delta^+(S,T) = \limsup_{l\to\infty} \frac{1}{l \log k} \limsupn \delta\left( \pi_{S,n}^{(l)} , \pi_{T,n}^{(l)} \right).
$$
\end{defn}
Recall that a \emph{pseudometric} on a set $X$ is a function $d: X \times X \to \R$ satisfying the following three conditions for all $x,y,z \in X$.
\begin{enumerate}[(i)]
    \item $d(x,y) \geq 0$, with equality if $x=y$. \ \ \ \ \ (nonnegativity)
    \label{pseudo-prop-1}
    \item $d(x,y) = d(y,x)$. \ \ \ \ \ \ \ \ \ \ \ \ \ \ \ \ \ \ \ \ \ \ \ \ \ \ \ \ (symmetry)
    \label{pseudo-prop-2}
    \item $d(x,z) \leq d(x,y) + d(y,z)$. \ \ \ \ \ \ \ \ \ \ \ \ \ \ \ \ (triangle inequality)
    \label{pseudo-prop-3}
\end{enumerate}
(A pseudometric is a \emph{metric}, or \emph{distance function}, on $X$ if it satisfies (\ref{pseudo-prop-1}) with ``if'' replaced by ``if and only if''.) The following fact must be known, but we do not know a reference at the time of this writing.

\begin{lem} \label{lem-delta-pseudometric}
For each positive integer $n$, the log-dispersion function $\delta$ is a pseudometric on $\Delta_n$.
\end{lem}

\newcommand{\proofOfLemDeltaPseudometric}{
    Let $n$ be a positive integer and let $\pi,\mu,\nu\in\Delta_n$. Since $\delta(\pi,\mu) = \log m$, where $m$ is a positive integer, $\delta(\pi,\mu) \geq 0$. Thus $\delta$ is nonnegative. If $\pi = \mu$, then it is easy to verify that the $n \times n$ identity matrix $\ma{I}_n$ testifies that $\delta(\pi,\mu) = 0$.

    To show that $\delta$ is symmetric, it suffices to prove that $\delta(\pi,\mu) \leq \delta(\mu,\pi)$. Let $m = 2^{\delta(\mu,\pi)}$. Then there exists an $n \times n$ nonnegative stochastic matrix $\ma{A} = (a_{ij})$ such that $\pi = \ma{A} \mu$ and $\ma{A}$ has at most $m$ nonzero entries in each row and column. Define the $n \times n$ matrix $\ma{A}' = (a'_{ij})$ for all $1 \leq i,j \leq n$ by
    $$
        a'_{ij} =
        \begin{cases}
            a_{ji} \dfrac{\mu(i)}{\pi(j)},         & \text{if $\pi(j)>0$} \\
            a_{ji} \dfrac{1}{\sum_{k=1}^n a_{jk}}, & \text{if $\pi(j)=0$}
        \end{cases}
    $$
    For all $1 \leq j \leq n$ such that $\pi(j) = 0$,
    $$
        \sum_{i=1}^n a'_{ij}
        = \sum_{i=1}^n a_{ji} \frac{1}{\sum_{k=1}^n a_{jk}}
        = 1.
    $$
    Since $\ma{A} \mu = \pi$, for all $1 \leq j \leq n$ such that $\pi(j) > 0$,
    $$
        \sum_{i=1}^n a'_{ij}
        = \sum_{i=1}^n a_{ji} \dfrac{\mu(i)}{\pi(j)}
        = \frac{1}{\pi(j)} \sum_{i=1}^n a_{ji} \mu(i)
        = \frac{1}{\pi(j)} \pi(j)
        = 1,
    $$
    so $\ma{A}'$ is stochastic. Since $\ma{A}$ is stochastic, for all $1 \leq i \leq n$,
    $$
        \sum_{j=1}^n a'_{ij} \pi(j)
        = \sum_{j=1}^n \left( a_{ji} \dfrac{\mu(i)}{\pi(j)} \right) \pi(j)
        = \sum_{j=1}^n a_{ji} \mu(i)
        = \mu(i),
    $$
    so $\ma{A}' \pi = \mu$. Since $a_{ji} = 0 \implies a'_{ij} = 0$, and $\ma{A}$ has at most $m$ nonzero entries in each row and column, $\ma{A}'$ has at most $m$ nonzero entries in each row and column as well. Thus $\delta(\pi,\mu) \leq \log m = \delta(\mu,\pi)$, so $\delta$ is symmetric.

    To see that $\delta$ satisfies the triangle inequality, let $m_1 = 2^{\delta(\pi,\mu)}$ and $m_2 = 2^{\delta(\mu,\nu)}$. It suffices to show that $\delta(\pi,\nu) \leq \log m_1 + \log m_2 = \log m_1 m_2$. There exist $n \times n$ nonnegative stochastic matrices $\ma{A}_1$ and $\ma{A}_2$ having no more than $m_1$ and $m_2$ nonzero entries in each row and column, respectively, satisfying $\ma{A}_1 \pi = \mu$ and $\ma{A}_2 \mu = \nu$. Let $\ma{A} = \ma{A}_2 \ma{A}_1$. Since the product of two stochastic matrices is stochastic, $\ma{A}$ is stochastic. Also, $\ma{A} \pi = \ma{A}_2 (\ma{A}_1 \pi) = \ma{A}_2 \mu = \nu$. Finally, since no row or column of $\ma{A}_1$ (resp. $\ma{A}_2$) contains more than $m_1$ (resp. $m_2$) nonzero entries, no row or column of $\ma{A}$ contains more than $m_1 m_2$ nonzero entries. Thus $\delta(\pi,\nu) \leq \log m_1 m_2$, so $\delta$ satisfies the triangle inequality.
}

\begin{journal}
    \begin{proof}
    \proofOfLemDeltaPseudometric
    \end{proof}
\end{journal}

It is easy to see that $S$ is not a metric on $\Delta_n$ for any $n \geq 2$. For example, if $\pi$ is any nonuniform probability measure on $\{1,\ldots,n\}$ and $\mu$ obtained from $\pi$ by permuting the values of $\pi$ nontrivially, then $\pi \neq \mu$ but $\delta(\pi,\mu) = 0$.

Lemma \ref{lem-delta-pseudometric} has the following immediate consequence.

\begin{cor} \label{cor-dispersion-pseudometric}
The normalized upper log-block dispersion function $\delta^+$ is a pseudometric on $\Sigma^\infty$.
\end{cor}

If $d$ is a pseudometric on a set $X$, then a function $f: X \to \R$ is \emph{$d$-contractive} if, for all $x,y \in X$,
$$
    |f(x) - f(y)| \leq d(x,y),
$$
i.e., the distance between $f(x)$ and $f(y)$ does not exceed the pseudodistance between $x$ and $y$.
\begin{journal}
    We prove the following lemma at the end of this section.
\end{journal}
\begin{conference-with-appendix}
    We use Schur concavity to prove the following lemma in the Technical Appendix.
\end{conference-with-appendix}

\begin{lem} \label{lem-entropy-contractive}
For each positive integer $n$, the Shannon entropy function \linebreak $H:\Delta_n \to [0,\log n]$ is $\delta$-contractive.
\end{lem}

The following useful fact follows easily from Lemma \ref{lem-entropy-contractive}.

\begin{theorem} \label{thm-dim-contractive}
Finite-state dimension and finite-state strong dimension are $\delta^+$-contractive. That is, for all $S,T\in\Sigma^\infty$,
$$
    |\dimfs(S) - \dimfs(T)| \leq \delta^+(S,T)
$$
and
$$
    |\Dimfs(S) - \Dimfs(T)| \leq \delta^+(S,T).
$$
\end{theorem}

In this paper, we only use the following special case of Theorem \ref{thm-dim-contractive}.

\begin{cor} \label{cor-dispersion-fsd-eq}
Let $S,T\in\Sigma^\infty$. If
$$
    \limsupn \delta\left( \pi_{S,n}^{(l)} , \pi_{T,n}^{(l)} \right) = o(l)
$$
as $l\to\infty$, then
$$
    \dimfs(S) = \dimfs(T)
$$
and
$$
    \Dimfs(S) = \Dimfs(T).
$$
\end{cor}

\newcommand{\SchurConcavity}{
    The proof of Lemma \ref{lem-entropy-contractive} uses Schur concavity \cite{Schu23,MarOlk79,Bhat97}, which we now review. We say that a vector $\ve{x} = (x_1,\ldots,x_n)\in\R^n$ is \emph{nonincreasing} if $x_1 \geq \ldots \geq x_n$. If $\ve{x},\ve{y}\in\R^n$ are nonincreasing, then we say that $\ve{x}$ \emph{majorizes} $\ve{y}$, and we write $\ve{x} \major \ve{y}$, if the following two conditions hold.
    \begin{enumerate}[(i)]
        \item $\sum_{i=1}^n x_i = \sum_{i=1}^n y_i$.
        \item For all $1 \leq t \leq n$, $\sum_{i=1}^t x_i \geq \sum_{i=1}^t y_i$.
    \end{enumerate}
    Given a vector $\ve{x}\in\R^n$ and a permutation $\pi$ of $\{1,\ldots,n\}$, write $\pi(\ve{x}) = (x_{\pi(1)} , \ldots , x_{\pi(n)})$. Call a set $D \subseteq \R^n$ \emph{symmetric} if $\pi(\ve{x}) \in D$ holds for every $\ve{x} \in D$ and every permutation $\pi$ of $\{1,\ldots,n\}$. For $D \subseteq \R^n$, a function $f: D \to \R$ is then \emph{symmetric} if $D$ is symmetric and $f(\ve{x}) = f(\pi(\ve{x}))$ holds for every $\ve{x} \in D$ and every permutation $\pi$ of $\{1,\ldots,n\}$.
    \begin{defn}
    Let $D \subseteq \R^n$ and $f: D \to \R$ be symmetric. Then $f$ is \emph{Schur-concave} if, for all $\ve{x},\ve{y} \in \R^n$,
    $$
        \ve{x} \major \ve{y} \implies f(\ve{x}) \leq f(\ve{y}).
    $$
    \end{defn}

    The set $\Delta_n$ of all probability measures on $\{1,\ldots,n\}$ can be regarded as the $(n-1)$-dimensional simplex
    $$
        \Delta_n = \left\{ \ve{p} \in [0,1]^n \ \left|\ \sum_{i=1}^n p_i = 1 \right. \right\} \subseteq \R^n.
    $$
    This set $\Delta_n$ is symmetric, as is the Shannon entropy function $H: \Delta_n \to [0,\log n]$. In fact, the following fundamental property of Shannon entropy is well known \cite{Bhat97}.

    \begin{lem} \label{lem-entropy-schur-concave}
    The Shannon entropy function $H: \Delta_n \to [0,\log n]$ is Schur-concave.
    \end{lem}
}

\newcommand{\proofLemEntopyContractive}{
    Fix a positive integer $n$, and let $\ve{p},\ve{q} \in \Delta_n$. By the symmetry of $\delta$ (established in Lemma \ref{lem-delta-pseudometric}), it suffices to prove that
    \begin{equation} \label{eq-lem-entropy-contractive}
        H(\ve{p}) \leq H(\ve{q}) + \delta(\ve{p},\ve{q}).
    \end{equation}
    Without loss of generality, assume that $\ve{p}$ and $\ve{q}$ are nonincreasing. Let $m$ be the positive integer such that $\delta(\ve{p},\ve{q}) = \log m$, and let $\ma{A} = (a_{ij})$ be an $n \times n$ matrix testifying to the value of $\delta(\ve{p},\ve{q})$. Define an $n \times n$ matrix $\ma{B} = (b_{ij})$ by
    $$
        b_{ij} =
        \begin{cases}
            1 & \text{if $(i-1)m < j \leq \min\{ im,n \}$;} \\
            0 & \text{otherwise.}
        \end{cases}
    $$
    That is, the first block of $m$ entries in the first row of $\ma{B}$ are 1's, the second block of $m$ entries in the second row of $\ma{B}$ are 1's, and so on, until the last $n - m \left( \ceil{\frac{n}{m}} - 1 \right)$ entries in the $\ceil{\frac{n}{m}}\Th$ row of $\ma{B}$ are 1's.

    Let $\ve{r} = \ma{B} \ve{p}$. Intuitively, $\ma{B}$ represents the ``worst-case'' matrix with no more than $m$ nonzero entries in each row and column, in the sense that it produces the vector with the lowest entropy. More formally, we show that $\ve{r}$ majorizes the vector $\ve{q}$, and thus $\ve{r}$ has entropy at most that of $\ve{q}$. However, since $\ma{B}$ is limited to $m$ nonzero entries in each row and column, it cannot redistribute the values in $\ve{p}$ by too much, so the entropy of $\ve{r}$ will be close to that of $\ve{p}$.

    Since $\ma{B}$ is stochastic (because each column contains exactly one 1) and $\ve{p} \in \Delta_n$, we have $\ve{r} \in \Delta_n$. Clearly, $\ve{r}$ is nonincreasing. For each $1 \leq j \leq n$, let $C_i = \{ j \ |\ a_{ij} > 0 \}$, noting that $|C_i| \leq m$. Then, for all $1 \leq t \leq n$,
    {
    $$
    \begin{array}{llllllllllllllllll}
        \sum\limits_{i=1}^t r_i
        &&=&&
        \sum\limits_{i=1}^t \sum\limits_{j=1}^n b_{ij} p_j
        &&=&&
        \sum\limits_{i=1}^t \sum\limits_{j=(i-1)m + 1}^{\min\{ im,n \}} p_j
        &&=&&
        \sum\limits_{i=1}^{\min\{ tm,n \}} p_i
        &&& \\ &&\geq&&
        \sum\limits_{j \in C_1 \cup \ldots \cup C_t} p_j
        &&\geq&&
        \sum\limits_{i=1}^t \sum\limits_{j \in C_i} a_{ij} p_j
        &&=&&
        \sum\limits_{i=1}^t \sum\limits_{j=1}^n a_{ij} p_j
        &&=&&
        \sum\limits_{i=1}^t q_i.
    \end{array}
    $$}
    (The first inequality holds because $\ve{p}$ is nonincreasing and each $|C_i| \leq m$. The second inequality holds because $\sum_{i=1}^n a_{ij} = 1$ holds for each $1 \leq j \leq n$, whence a single $p_j$'s appearances in various $C_i$'s collectively contribute at most $p_j$ to the sum on the right.) This shows that $\ve{r} \major \ve{q}$, whence Lemma \ref{lem-entropy-schur-concave} tells us that $H(\ve{r}) \leq H(\ve{q})$. It follows by Jensen's inequality and the (ordinary) concavity of the logarithm that
    \begin{eqnarray*}
        H(\ve{p})
        &\leq&
        H(\ve{p}) + \log \sum_{i=1}^n p_i \frac{1}{p_i} r_{\ceil{\frac{i}{m}}} - \sum_{i=1}^n p_i \log \left( \frac{1}{p_i} r_{\ceil{\frac{i}{m}}} \right)
        \\ &=&
        H(\ve{p}) + \log \sum_{i=1}^n r_{\ceil{\frac{i}{m}}} - \sum_{i=1}^n p_i \log \left( \frac{1}{p_i} r_{\ceil{\frac{i}{m}}} \right)
        \\ &\leq&
        H(\ve{p}) + \log m - \sum_{i=1}^n p_i \log \left( \frac{1}{p_i} r_{\ceil{\frac{i}{m}}} \right)
        \\ &=&
        \sum_{i=1}^n p_i \log \frac{1}{r_{\ceil{\frac{i}{m}}}} + \log m
        \\ &=&
        \sum_{i=1}^n r_i \log \frac{1}{r_i} + \log m
        \\ &=&
        H(\ve{r}) + \delta(\ve{p},\ve{q})
        \\ &\leq&
        H(\ve{q}) + \delta(\ve{p},\ve{q}),
    \end{eqnarray*}
    i.e., (\ref{eq-lem-entropy-contractive}) holds.
}

\begin{journal}
    \SchurConcavity

    We now use Lemma \ref{lem-entropy-schur-concave} to prove Lemma \ref{lem-entropy-contractive}.

    \begin{proof}[Proof of Lemma \ref{lem-entropy-contractive}]
    \proofLemEntopyContractive
    \end{proof}
\end{journal}

\newcommand{\dimfsk}{\mathrm{dim}_\mathrm{FS}^{(k)}}
\newcommand{\Dimfsk}{\mathrm{Dim}_\mathrm{FS}^{(k)}}
\section{Finite-State Dimension and Real Arithmetic}
Our main theorem concerns real numbers rather than sequences, so the following notation is convenient. For each real number $\alpha$ and each integer $k \geq 2$, write
$$
    \dimfsk(\alpha) = \dimfs(S)
$$
and
$$
    \Dimfsk(\alpha) = \Dimfs(S),
$$
where $S$ is a base-$k$ expansion of $\alpha - \floor{\alpha}$. Note that this notation is well-defined, because a real number $\alpha$ has two base-$k$ expansions if and only if it is a $k$-adic rational, in which case both expansions are eventually periodic and hence have finite-state strong dimension 0. It is routine to verify the following.

\begin{obs} \label{obs-addition}
For every integer $k \geq 2$, every positive integer $m$, and every real number $\alpha$,
$$
    \dimfsk(m + \alpha) = \dimfsk(-\alpha) = \dimfsk(\alpha)
$$
and
$$
    \Dimfsk(m + \alpha) = \Dimfsk(-\alpha) = \Dimfsk(\alpha).
$$
\end{obs}

The following lemma contains most of the technical content of our main theorem.

\begin{lem}[main lemma] \label{lem-main}
For every integer $k \geq 2$, every positive integer $m$, and every real number $\alpha \geq 0$,
$$
    \dimfsk(m \alpha) = \dimfsk(\alpha)
$$
and
$$
    \Dimfsk(m \alpha) = \Dimfsk(\alpha).
$$
\end{lem}

\newcommand{\proofMainLemma}{
    Let $k$, $m$, and $\alpha$ be as given, let $S,T\in\Sigma^\infty$ be the base-$k$ expansions of $\alpha - \floor{\alpha}$, $m \alpha - \floor{m \alpha}$, respectively, and write
    $$
        \pi_{\alpha,n}^{(l)} = \pi_{S,n}^{(l)} \ \ \ \ , \ \ \ \  \pi_{m \alpha,n}^{(l)} = \pi_{T,n}^{(l)}
    $$
    for each $l,n\in\Z^+$. By Corollary \ref{cor-dispersion-fsd-eq}, it suffices to show that
    \begin{equation} \label{eq-lem-main}
        \limsupn \delta\left( \pi_{\alpha,n}^{(l)} , \pi_{m \alpha,n}^{(l)} \right) = o(l)
    \end{equation}
    as $l\to\infty$.

    Let $r = \floor{\log_k m}$, let
    $$
        m = \sum_{i=0}^r m_i k^i
    $$
    be the base-$k$ expansion of $m$, and let
    $$
        s = \sum_{i=0}^r m_i.
    $$
    The first thing to note is that, in base $k$, $m \alpha - \floor{m \alpha}$ is the sum, modulo 1, of $s$ copies of $\alpha - \floor{\alpha}$, with $m_i$ of these copies shifted $i$ symbols to the left, for each $0 \leq i \leq r$.

    For each $l\in\Z^+$ and $j\in\N$, let
    \begin{eqnarray*}
        u_j^{(l)} &=& S[jl \twodots (j+1)l-1], \\
        v_j^{(l)} &=& T[jl \twodots (j+1)l-1]
    \end{eqnarray*}
    be the $j\Th$ $l$-symbol blocks of $\alpha - \floor{\alpha}$, $m \alpha - \floor{m \alpha}$, respectively. If we let
    $$
        \tau_j^{(l)} = \sum_{i=0}^r m_i \sum_{t=(j+1)l}^\infty S[t+i] k^{-(t+1)}
    $$
    be the sum of the tails of the above-mentioned $s$ copies of $\alpha - \floor{\alpha}$ lying to the right of the $j\Th$ $l$-symbol block, then the block $v_j^{(l)}$ of $m \alpha - \floor{m \alpha}$ is completely determined by $u_j^{(l)}$, the ``carry''
    $$
        c_j^{(l)} = \floor{k^{(j+1)l} \tau_j^{(l)}},
    $$
    and the longest string of symbols shifted from the right, which is the string $u_{j+1}^{(l)}[0 \twodots r-1]$. To be more explicit, note that
    $$
        0 \leq c_j^{(l)} \leq k^{(j+1)l} \tau_j^{(l)} \leq k^{(j+1)l} \sum_{i=0}^r m_i \sum_{t=(j+1)l}^\infty (k-1) k^{-(t+1)} = s;
    $$
    define the ``advice''
    $$
        h_j^{(l)} = \left( c_j^{(l)} , u_j^{(l)}[0 \twodots r-1] \right) \in \{0,\ldots,s\} \times \Sigma^r;
    $$
    and define the function
    $$
        f^{(l)}: \Sigma^l \times \{0,\ldots,s\} \times \Sigma^r \to \Sigma^l
    $$
    by letting $f^{(l)}(x,c,z)$ be the base-$k$ expansion of the integer
    $$
        m n_x^{(k)} + c + \sum_{i=0}^r m_i \sum_{t=0}^{i-1} z[t] k^t \text{ \bf mod } k^l,
    $$
    where $n_x^{(k)}$ is the nonnegative integer of which $x$ is a base-$k$ expansion, possibly with leading 0's. (Intuitively, the three terms here are the ``block product'', the ``carry'', and the ``shift'', respectively.) Then, for all integers $l>0$ and $j \geq 0$,
    $$
        v_j^{(l)} = f^{(l)}(u_j^{(l)} , h_j^{(l)}).
    $$

    For positive integers $l$ and $n$, define the $k^l \times k^l$ matrix $\ma{A}^{(l,n)} = \left( a_{y,x}^{(l,n)} \right)$ by
    $$
    a_{y,x}^{(l,n)} =
    \begin{cases}
        \dfrac{ \left| \left\{ \ j<n \ \left|\ u_j^{(l)} = x \text{ and } f^{(l)} \left( x,h_j^{(l)}(j) \right) = y \right. \right\} \right| }{n \pi_{\alpha,n}^{(l)}(x)} & \text{if $\pi_{\alpha,n}^{(l)}(x) > 0$} \\
        1 & \text{if $\pi_{\alpha,n}^{(l)}(x) = 0$ and $x=y$} \\
        0 & \text{otherwise} \\
    \end{cases}
    $$
    for all $x,y\in\Sigma^l$. It is routine to verify that
    $$
        \sum_{y\in\Sigma^l} a_{y,x}^{(l,n)} = 1
    $$
    for all $x\in\Sigma^l$, i.e., $\ma{A}^{(l,n)}$ is stochastic, and that
    $$
        \sum_{x\in\Sigma^l} a_{y,x}^{(l,n)} \pi_{\alpha,n}^{(l)}(x) = \pi_{m \alpha,n}^{(l)}(y)
    $$
    for all $y\in\Sigma^l$, i.e., $\ma{A}^{(l,n)} \pi_{\alpha,n}^{(l)} = \pi_{m \alpha,n}^{(l)}$. We complete the proof by bounding the number of nonzero entries in each row and column of $\ma{A}^{(l,n)}$.

    Fix a column $x$ of $\ma{A}^{(l,n)}$. If $\pi_{\alpha,n}^{(l)}(x) = 0$, then there is exactly one nonzero entry in column $x$ of $\ma{A}^{(l,n)}$. If $\pi_{\alpha,n}^{(l)}(x) > 0$, then the number of nonzero entries in column $x$ is bounded by
    $$
        |\{0,\ldots,s\} \times \Sigma^r| = (s+1)k^r \leq (s+1)m.
    $$
    Hence there are at most $(s+1)m$ nonzero entries in column $x$ of $\ma{A}^{(l,n)}$.

    Fix a row $y$ of $\ma{A}^{(l,n)}$. Let $g$ be the greatest common divisor of $m$ and $k^l$. Note that, for all $n_1,n_2 \in \Z^+$,
    \begin{eqnarray*}
        m n_1 \equiv m n_2 \mod k^l
        &\implies&
        k \ |\ m (n_2-n_1)
        \\ &\implies&
        \left. \frac{k^l}{g} \ \right|\ \frac{m}{g} (n_2-n_1)
        \\ &\implies&
        \left. \frac{k^l}{g} \ \right|\ n_2-n_1
        \\ &\implies&
        n_1 \equiv n_2 \mod \frac{k^l}{g}.
    \end{eqnarray*}
    This implies that each string $y\in\Sigma^l$ has at most $g$ preimages $x$ under the mapping that takes $x$ to the base-$k$ expansion of $m n_x^{(l)} \text{ \bf mod } k^l$. This, in turn, implies that there are at most $g |\{0,\ldots,s\} \times \Sigma^r| \leq g(s+1)m$ nonzero entries in row $y$ of $\ma{A}^{(l,n)}$.

    We have shown that, for each $l,n\in\Z^+$, the matrix $\ma{A}^{(l,n)}$ testifies that
    $$
        \delta\left( \pi_{\alpha,n}^{(l)} , \pi_{m \alpha,n}^{(l)} \right)
        \leq
        \log(g(s+1)m) \leq \log(m^2(s+1)).
    $$
    Since this bound does not depend on $l$ or $n$, this proves (\ref{eq-lem-main}).
}

\begin{proof}
\proofMainLemma
\qedllncs
\end{proof}

We now prove that addition and multiplication by nonzero rationals preserve finite-state dimension and finite-state strong dimension.

\begin{theorem}[main theorem] \label{thm-main}
For every integer $k \geq 2$, every nonzero rational number $q$, and every real number $\alpha$,
$$
    \dimfsk(q + \alpha) = \dimfsk(q \alpha) = \dimfsk(\alpha)
$$
and
$$
    \Dimfsk(q + \alpha) = \Dimfsk(q \alpha) = \Dimfsk(\alpha).
$$
\end{theorem}

\begin{proof}
Let $k$, $q$, and $\alpha$ be as given, and write $q = \frac{a}{b}$, where $a$ and $b$ are integers with $a \neq 0$ and $b > 0$. By Observation \ref{obs-addition} and Lemma \ref{lem-main},
$$
\begin{array}{llllllllllllllllll}
    \dimfsk(q \alpha)
    &&=&&
    \dimfsk\left( \dfrac{|a|}{b} \alpha \right)
    &&=&&
    \dimfsk\left( b \dfrac{|a|}{b} \alpha \right)
    &&& \\ &&=&&
    \dimfsk(|a| \alpha)
    &&=&&
    \dimfsk(\alpha),
\end{array}
$$
and
$$
\begin{array}{llllllllllllllllll}
    \dimfsk(q + \alpha)
    &&=&&
    \dimfsk\left( \dfrac{a}{b} + \alpha \right)
    &&=&&
    \dimfsk\left( \dfrac{a + b \alpha}{b} \right)
    &&& \\ &&=&&
    \dimfsk\left( b \dfrac{a + b \alpha}{b} \right)
    &&=&&
    \dimfsk(a + b \alpha)
    &&& \\ &&=&&
    \dimfsk(b \alpha)
    &&=&&
    \dimfsk(\alpha).
\end{array}
$$
Similarly, $\Dimfsk(q \alpha) = \Dimfsk(\alpha)$, and $\Dimfsk(q + \alpha) = \Dimfsk(\alpha)$.
\qedllncs
\end{proof}

Finally, we note that Theorem \ref{thm-main} gives a new proof of the following classical theorem.

\begin{cor}(Wall \cite{Wall49})
Let $k \geq 2$. For every nonzero rational number $q$ and every real number $\alpha$ that is normal base $k$, the sum $q + \alpha$ and the product $q \alpha$ are also normal base $k$.
\end{cor}

\begin{ack}
The authors thank Philippe Moser and Arindam Chatterjee for useful discussions.
\end{ack}

\bibliography{dim,rbm,main,random,dimrelated}

\begin{thebibliography}{10}

\bibitem{Agaf68}
V.~N. Agafonov.
\newblock Normal sequences and finite automata.
\newblock {\em Soviet Mathematics Doklady}, 9:324--325, 1968.

\bibitem{Athreya:ESDAICC}
K.~B. Athreya, J.~M. Hitchcock, J.~H. Lutz, and E.~Mayordomo.
\newblock Effective strong dimension, algorithmic information, and
  computational complexity.
\newblock {\em SIAM Journal on Computing}, 2004.
\newblock To appear. Preliminary version appeared in {\em Proceedings of the
  21st International Symposium on Theoretical Aspects of Computer Science},
  pages 632-643.

\bibitem{Bhat97}
R.~Bhatia.
\newblock {\em Matrix Analysis}.
\newblock Springer, 1997.

\bibitem{Bore09}
\'E. Borel.
\newblock Sur les probabilit\'es d\'enombrables et leurs applications
  arithm\'etiques.
\newblock {\em Rend. Circ. Mat. Palermo}, 27:247--271, 1909.

\bibitem{BorBai04}
J.~Borwein and D.~Bailey.
\newblock {\em Mathematics by Experiment: Plausible Reasoning in the $21^{\rm
  st}$ Century}.
\newblock A. K. Peters, Ltd., Natick, MA, 2004.

\bibitem{Bourke:ERFSD}
C.~Bourke, J.~M. Hitchcock, and N.~V. Vinodchandran.
\newblock Entropy rates and finite-state dimension.
\newblock {\em Theoretical Computer Science}, 2005.
\newblock To appear.

\bibitem{Cass59}
J.~W.~S. Cassels.
\newblock On a problem of {Steinhaus} about normal numbers.
\newblock {\em Colloquium Mathematicum}, 7:95--101, 1959.

\bibitem{Cham33}
D.~G. Champernowne.
\newblock Construction of decimals normal in the scale of ten.
\newblock {\em J. London Math. Soc.}, 2(8):254--260, 1933.

\bibitem{Dai:FSD}
J.~J. Dai, J.~I. Lathrop, J.~H. Lutz, and E.~Mayordomo.
\newblock Finite-state dimension.
\newblock {\em Theoretical Computer Science}, 310:1--33, 2004.
\newblock Preliminary version appeared in \emph{Proceedings of the 28th
  International Colloquium on Automata, Languages, and Programming}, pages
  1028--1039, 2001.

\bibitem{DajKra02}
K.~Dajani and C.~Kraaikamp.
\newblock {\em Ergodic Theory of Numbers}.
\newblock The Mathematical Association of America, 2002.

\bibitem{Edg03}
G.~A. Edgar.
\newblock {\em Classics on Fractals}.
\newblock Westview Press, Oxford, U.K., 2004.

\bibitem{Falc90}
K.~Falconer.
\newblock {\em Fractal Geometry: Mathematical Foundations and Applications}.
\newblock John Wiley \& Sons, 1990.

\bibitem{Harm03}
G.~Harman.
\newblock One hundred years of normal numbers.
\newblock In {\em M. A. Bennett, B. C. Berndt, N. Boston, H. G. Diamond, A. J.
  Hildebrand, and W. Philip (eds.), Surveys in Number Theory: Papers from the
  Millennial Conference on Number Theory}, pages 57--74, 2003.

\bibitem{Haus19}
F.~Hausdorff.
\newblock Dimension und {\"a}usseres {M}ass.
\newblock {\em Mathematische Annalen}, 79:157--179, 1919.
\newblock English version appears in \cite{Edg03}, pp. 75-99.

\bibitem{Hitchcock:FDLLU}
J.~M. Hitchcock.
\newblock Fractal dimension and logarithmic loss unpredictability.
\newblock {\em Theoretical Computer Science}, 304(1--3):431--441, 2003.

\bibitem{Kam73}
T.~Kamae.
\newblock Subsequences of normal sequences.
\newblock {\em Israel Journal of Mathematics}, 16:121--149, 1973.

\bibitem{KamWei75}
T.~Kamae and B.~Weiss.
\newblock Normal numbers and selection rules.
\newblock {\em Israel Journal of Mathematics}, 21:101--110, 1975.

\bibitem{KuiNie74}
L.~Kuipers and H.~Niederreiter.
\newblock {\em Uniform Distribution of Sequences}.
\newblock Wiley-Interscience, 1974.

\bibitem{LiVi97}
M.~Li and P.~M.~B. Vit\'{a}nyi.
\newblock {\em An Introduction to Kolmogorov Complexity and its Applications}.
\newblock Springer-Verlag, Berlin, 1997.
\newblock Second Edition.

\bibitem{Lutz:DCC}
J.~H. Lutz.
\newblock Dimension in complexity classes.
\newblock {\em SIAM Journal on Computing}, 32:1236--1259, 2003.
\newblock Preliminary version appeared in {\em Proceedings of the Fifteenth
  Annual IEEE Conference on Computational Complexity}, pages 158--169, 2000.

\bibitem{Lutz:DISS}
J.~H. Lutz.
\newblock The dimensions of individual strings and sequences.
\newblock {\em Information and Computation}, 187:49--79, 2003.
\newblock Preliminary version appeared in \emph{Proceedings of the 27th
  International Colloquium on Automata, Languages, and Programming}, pages
  902--913, 2000.

\bibitem{MarOlk79}
A.~W. Marshall and I.~Olkin.
\newblock {\em Inequalities: Theory of Majorization and Its Applications}.
\newblock Academic Press, New York, 1979.

\bibitem{MerRei03}
Wolfgang Merkle and Jan Reimann.
\newblock On selection functions that do not preserve normality.
\newblock In Branislav Rovan and Peter Vojt{\'a}s, editors, {\em MFCS}, volume
  2747 of {\em Lecture Notes in Computer Science}, pages 602--611, Bratislava,
  Slovakia, 2003. Springer.

\bibitem{Nive56}
I.~Niven.
\newblock {\em Irrational Numbers}.
\newblock Wiley, 1956.

\bibitem{Schm60}
W.~Schmidt.
\newblock On normal numbers.
\newblock {\em Pacific Journal of Mathematics}, 10:661--672, 1960.

\bibitem{SchSti72}
C.~P. Schnorr and H.~Stimm.
\newblock Endliche {Automaten} und {Zufallsfolgen}.
\newblock {\em Acta Informatica}, 1:345--359, 1972.

\bibitem{Schu23}
I.~Schur.
\newblock {Uber eine Klasse von Mittelbildungen mit Anwendungen auf die
  Determinantentheorie}.
\newblock {\em Math. Ges.}, 22:9--20, 1923.

\bibitem{Sull84}
D.~Sullivan.
\newblock Entropy, {Hausdorff} measures old and new, and limit sets of
  geometrically finite {Kleinian} groups.
\newblock {\em Acta Mathematica}, 153:259--277, 1984.

\bibitem{Tricot82}
C.~Tricot.
\newblock Two definitions of fractional dimension.
\newblock {\em Mathematical Proceedings of the Cambridge Philosophical
  Society}, 91:57--74, 1982.

\bibitem{Wall49}
D.~D. Wall.
\newblock {\em Normal Numbers}.
\newblock PhD thesis, University of California, Berkeley, California, USA,
  1949.

\bibitem{Weis99}
B.~Weiss.
\newblock {\em Single Orbit Dynamics}.
\newblock American Mathematical Society, Providence, RI, 2000.

\end{thebibliography}
\bibliographystyle{plain}

\begin{conference-with-appendix}

    \newpage
    \section{Technical Appendix}
    This appendix contains proofs of Lemmas \ref{lem-delta-pseudometric} and \ref{lem-entropy-contractive}.

    \begin{proof}[of Lemma \ref{lem-delta-pseudometric}]
    \proofOfLemDeltaPseudometric
    \qedllncs
    \end{proof}

    \SchurConcavity

    We now use Lemma \ref{lem-entropy-schur-concave} to prove Lemma \ref{lem-entropy-contractive}.

    \begin{proof}[of Lemma \ref{lem-entropy-contractive}]
    \proofLemEntopyContractive
    \qedllncs
    \end{proof}

\end{conference-with-appendix}

\end{document}